\documentclass[preprint,superscriptaddress]{revtex4-1}

\usepackage{graphicx}
\usepackage{dcolumn}
\usepackage{bm}
\usepackage{amsmath}
\usepackage{amssymb}
\usepackage{latexsym}
\usepackage{hyperref} 
\usepackage{placeins}

\usepackage{color,soul} 

\newcommand{\rfig}[1]{Fig.~\ref{#1}}
\newcommand{\rfigs}[1]{Figs.~\ref{#1}}
\newcommand{\rtab}[1]{Table.~\ref{#1}}

\newcommand{\req}[1]{Eq.~(\ref{#1})}
\newcommand{\reqs}[1]{Eqs.~(\ref{#1})}

\makeindex

\begin{document}

\title{Homointerface planar Josephson junction based on inverse proximity effect}

\author{Juewen Fan}
\author{Bingyan Jiang}
\author{Jiaji Zhao}
\author{Ran Bi}
\affiliation{State Key Laboratory for Artificial Microstructure and Mesoscopic Physics, Frontiers Science Center for Nano-optoelectronics, Peking University, Beijing 100871, China}
\author{Jiadong Zhou}
\affiliation{Key Lab of Advanced Optoelectronic Quantum Architecture and Measurement (Ministry of Education), Beijing Key Lab of Nanophotonics \& Ultrafine Optoelectronic Systems, and School of Physics, Beijing Institute of Technology, Beijing 100081, China}
\author{Zheng Liu}
\affiliation{School of Materials Science and Engineering, Nanyang Technological University, Singapore 639798, Singapore}
\author{Ning Kang}
\affiliation{Key Laboratory for the Physics and Chemistry of Nanodevices and Department of Electronics, Peking University, Beijing 100871, China}
\author{Fanming Qu}
\author{Li Lu}
\affiliation{Beijing National Laboratory for Condensed Matter Physics, Institute of Physics, Chinese Academy of Sciences, Beijing 100190, China}
\author{Xiaosong Wu}
\email{xswu@pku.edu.cn}
\affiliation{State Key Laboratory for Artificial Microstructure and Mesoscopic Physics, Frontiers Science Center for Nano-optoelectronics, Peking University, Beijing 100871, China}
\affiliation{Collaborative Innovation Center of Quantum Matter, Beijing 100871, China}
\affiliation{Shenzhen Institute for Quantum Science and Engineering, Southern University of Science and Technology, Shenzhen 518055, China}

\begin{abstract}
The quality of a superconductor{\textendash}normal metal{\textendash}superconductor Josephson junction (JJ) depends crucially on the transparency of the superconductor{\textendash}normal metal (S/N) interface. We demonstrate a technique for fabricating planar JJs with perfect S/N interfaces. The technique utilizes a strong inverse proximity effect discovered in Al/V$_5$S$_8$ bilayers, by which the Al layer is driven into the resistive state. The highly transparent S/N homointerface and the peculiar normal metal enable the flow of Josephson supercurrent across a 2.9 $\mu$m long weak link. Moreover, our JJ exhibits a giant critical current and a large product of the critical current and the normal state resistance.
\end{abstract}

\maketitle

\section{INTRODUCTION}

A Josephson junction (JJ) consists of two superconductors coupled through a weak link and is the fundamental element in a variety of superconducting electronics\cite{Hayakawa2004,Clarke2008,Hamilton2000}. Depending on the weak link, there are different types of JJs\cite{Golubov2004}, e.g., superconductor{\textendash}insulator{\textendash}superconductor (SIS)\cite{Dolan1988}, superconductor{\textendash}normal metal{\textendash}superconductor (SNS)\cite{Dubos2000,Chiodi2012}, superconductor{\textendash}constriction {\textendash}superconductor (ScS)\cite{Zaitsev1998Apr}, superconductor{\textendash}ferromagnet{\textendash}superconductor (SFS)\cite{Bolginov2018Mar,Singh2016Oct}, and superconductor{\textendash}two-dimensional
electron gas{\textendash}superconductor\cite{Shabani2016,Fornieri2019}. SNS JJs have a negligible inherent capacitance. Being overdamped, their current{\textendash}voltage ($I${\textendash}$V$) characteristics can be made, in principle, non-hysteretic, which is desired in high-frequency applications\cite{Belogolovskii2017}. Moreover, they have potentially higher $I_\mathrm{c}R_\mathrm{N}$ value\cite{Kulik1975}, which is the figure of merit in many applications\cite{Belogolovskii2017,Yu2006}. Here, $I_\mathrm{c}$ is the critical current and $R_\mathrm{N}$ is the normal state resistance. Recently, the interest in SNS JJs has been intensified, as it has been proposed that such junctions, when N is topologically nontrivial, can be used in topological quantum computing\cite{Fu2008}. However, the characteristics of SNS devices are strongly affected by the superconductor{\textendash}normal metal interface, which poses a challenge in device fabrication. It is also known that disorders at the interface lead to the soft-gap problem\cite{Takei2013Apr}. A large interface resistance causes a fast decay of Josephson supercurrent with
temperature or field\cite{Chiodi2012,Hammer2007}. Various techniques have been employed to achieve transparent and consistent interfaces, e.g., shadow deposition\cite{Dolan1988,Dubos2000}, \textit{in situ} epitaxial growth\cite{Krogstrup2015Apr,Shabani2016,Fornieri2019}, focused ion beam deposition\cite{Chiodi2012}, and trilayer technique\cite{Bolginov2018Mar}.

Here, we demonstrate a technique for constructing SNS JJs utilizing the inverse proximity effect (IPE). We observed a strong suppression of the superconducting transition temperature in a 31 nm aluminum film deposited on a 10 nm novel V$_5$S$_8$ superlattice film. Using this non-superconducting Al/V$_5$S$_8$ bilayer as the weak link, aluminum SNS JJs with fully transparent superconductor{\textendash}normal metal homointerfaces were fabricated. Such junctions exhibit high critical currents and large $I_\mathrm{c}R_\mathrm{N}$ values. 

\section{EXPERIMENTAL METHODS}

\begin{figure}[htbp]
	\begin{center}
		\includegraphics[width=0.8\columnwidth]{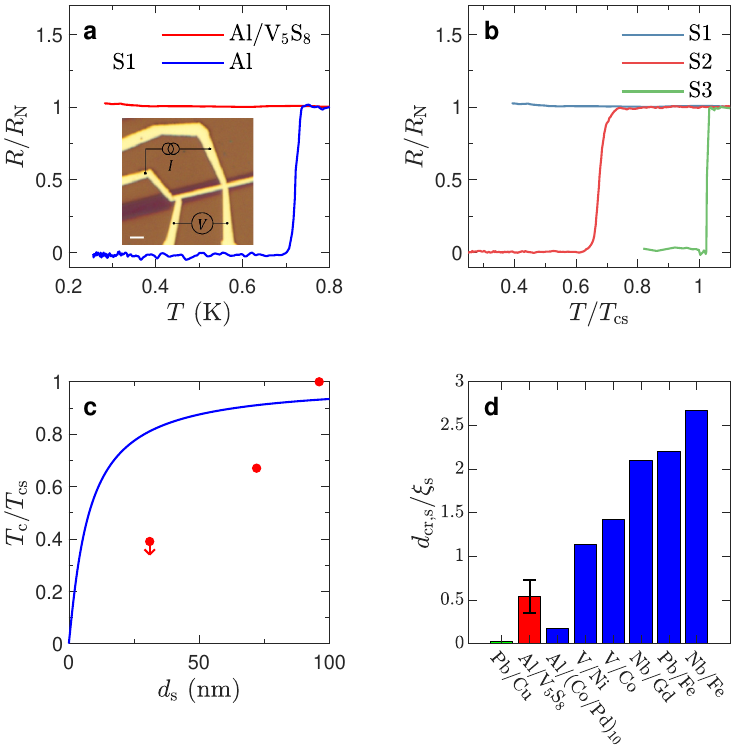}
		\caption{IPE in the bilayer of Al/V$_5$S$_8$. (a) Temperature dependence of the resistance for a 31-nm-thick Al film on V$_5$S$_8$ (device S1) and SiO$_2$. The resistance $R$ is normalized by $R_\mathrm{N}$. The inset is an illustration of the four-probe measurement of device S1. The scale bar represents 2 $\mu$m. (b) Normalized $R$ vs $T$ for device S1, S2, and S3. The Al thicknesses in these devices are 31, 72, and 96 nm, respectively. $T$ is normalized by $T_{\mathrm{cs}}$, the transition temperature of the corresponding Al film on the SiO$_2$ substrate. (c) Normalized transition temperature $T_{\mathrm{c}}/T_{\mathrm{cs}}$ vs thickness $d_{\mathrm{s}}$ of Al in the Al/V$_5$S$_8$ bilayer system. Dots with arrow denote the upper limit of the critical temperature as the device is not superconducting down to the lowest temperature of this study. The solid line represents the result of the Werthamer theory on the proximity effect in the S/N bilayer. (d) Comparison of $d_{\mathrm{cr,s}}/\xi_{\mathrm{s}}$ in Al on V$_5$S$_8$ with those in Pb/Cu\cite{Hilsch1962}, Al/(Co/Pd)$_{10}$\cite{Xia2009}, V/Ni, V/Co\cite{Aarts1997}, Nb/Gd\cite{Strunk1994}, Pb/Fe\cite{Lazar2000}, and Nb/Fe\cite{Muehge1997}.}
		\label{f1}
	\end{center}
\end{figure}

V$_5$S$_8$ superlattice films used in this study were grown by a chemical vapor deposition method on $\mathrm{SiO_2}$ substrates\cite{Zhou2021}. Devices were patterned using standard e-beam lithography, followed by e-beam deposition of a 2 nm titanium adhesion layer and the aluminum layer of desired thickness. We prepared two types of devices, that is, Al/V$_5$S$_8$ bilayer [see the inset of \rfig{f1}(a)] and JJ made from an Al and V$_5$S$_8$ crossbar [see \rfigs{f2}(a) and 2(b)]. Parameters for all devices are summarized in \rtab{t1}. Devices were loaded into an Oxford $^3$He cryostat with a base temperature of about 250 mK. Low-temperature electrical measurements, with multiple-stage low-pass filtering, were carried out using a lock-in amplifier.

\begin{table}[htbp]
\centering
        \caption{Device number, Al thickness $d_\mathrm{s}$, V$_5$S$_8$ thickness $d_\mathrm{n}$, junction length $L$, junction width $W$, superconducting transition temperature $T_{\mathrm{c}}$, sheet resistance $R_\mathrm{s}$, and mean free path $l_\mathrm{e}$ for the samples of this study. Due to the different length-width ratios of Al and V$_5$S$_8$ in the Al/V$_5$S$_8$ bilayer, for bilayer devices, S1{\textendash}3, $R_\mathrm{s}$ and $l_\mathrm{e}$ are not available because of non-negligible contribution from V$_5$S$_8$, which is much wider than the Al bar, seen in the inset of \rfig{f1}(a).}
        \begin{center}
        \begin{tabular}{p{0.8cm}p{3.8cm}p{1.5cm}p{1.5cm}p{1.5cm}p{1.5cm}p{1.5cm}p{1.7cm}p{1.4cm}}
            \hline \hline
            No. & Sample type & $d_\mathrm{s}$ (nm) & $d_\mathrm{n}$ (nm) & $L$ ($\mu$m) & $W$ ($\mu$m) & $T_{\mathrm{c}}$ (K) & $R_\mathrm{s}$ ($\Omega$/$\square$) & $l_\mathrm{e}$ (nm)\\
            \hline
            S1 & Al/V$_5$S$_8$ bilayer & 31 & 10 & & & $<$0.28 & & \\
            S2 & Al/V$_5$S$_8$ bilayer & 72 & 10 & & & 0.70 & & \\
            S3 & Al/V$_5$S$_8$ bilayer & 96 & 10 & & & 0.95 & & \\
            S4 & Al{\textendash}(Al/V$_5$S$_8$){\textendash}Al JJ & 31 & 10 & 1.1 & 2.7 & 0.73 & 1.35 & 6.2\\
            S5 & Al{\textendash}(Al/V$_5$S$_8$){\textendash}Al JJ & 31 & 10 & 2.9 & 1.6 & 0.31 & 0.88 & 9.5 \\
            S6 & Al{\textendash}(Al/V$_5$S$_8$){\textendash}Al JJ & 31 & 10 & 0.9 & 2.8 & 0.77 & 1.24 & 6.7\\
            \hline
        \end{tabular}  
        \end{center}
        \label{t1}
\end{table}

The compound of V$_5$S$_8$ known in the literature is VS$_2$ layers self-intercalated with vanadium\cite{Kawada1975}. It becomes an antiferromagnet below 32 K\cite{Nozaki1978}. In stark contrast, V$_5$S$_8$ used in this study has a unique superlattice structure consisting of VS$_2$ layers intercalated with V$_2$S$_2$ atomic chains\cite{Zhou2021}. It shows no indication of ferromagnetism. See supplementary material for more details. More interestingly, the new V$_5$S$_8$ displays an exotic in-plane Hall effect that has not been reported before. This effect results from an out-of-plane Berry curvature induced by the in-plane magnetic field, enabled by a peculiar anisotropic spin{\textendash}orbit coupling. For simplicity, we still use the chemical formula of V$_5$S$_8$ in this letter to refer to the new material.

\section{RESULTS AND DISCUSSION}

The bilayer device S1, Al(31)/V$_5$S$_8$(10), is illustrated in the inset of \rfig{f1}(a). The number in the parenthesis denotes the thickness in nanometer. The superconducting transition temperature of a 31 nm Al film on the SiO$_2$ substrate, defined by the midpoint of the resistance transition, is 0.72 K. The critical temperature is relatively low. This could be due to a relatively low crystalline quality, or even possible residue of vanadium compounds, introduced in the chemical vapor deposition growth that have magnetic local moments. Nevertheless, this will not affect the following analysis. Surprisingly, S1 remains in the normal state down to 0.28 K. Not even a slight depression of the resistance was observed, indicating that 31 nm Al is driven into a resistive state by 10 nm V$_5$S$_8$. In contrast, superconductor films on many other layered transition metal dichalcogenides stay in the superconducting state\cite{Wu2019,Lyu2018Oct,Huang2018Jul,Luepke2020May,Trainer2020Mar,Ghatak2018}. To get an idea of the strength of the IPE, we fabricated more bilayer devices, in which the thickness of the Al film is varied, while maintaining the same V$_5$S$_8$ thickness as in device S1. As shown in \rfig{f1}(b), the superconducting transition temperature of bilayer $T_{\mathrm{c}}$ gradually recovers to $T_{\mathrm{cs}}$ with increasing Al thickness. Here, $T_{\mathrm{cs}}$ is the critical temperature of the Al layer on $\mathrm{SiO_2}$ that was deposited along with the bilayer. When the Al film is 96 nm, $T_{\mathrm{c}}$ is equal to $T_{\mathrm{cs}}$. We compare our experimental data with the classical theory describing the proximity effect in the S/N bilayer\cite{Werthamer1963,Hauser1964,Nagel1994}. The detailed calculation can be found in the supplementary material. As shown in \rfig{f1}(c), the experimental data are below the theoretical calculation. Apparently, the IPE induced by V$_5$S$_8$ on the Al film is unexpectedly strong.

When the superconductor layer in an S/N bilayer is thin, the system is non-superconducting due to the IPE. As the thickness of the superconductor layer increases to a critical value $d_{\mathrm{cr,s}}$, the system turns into the superconducting state. Therefore, $d_{\mathrm{cr,s}}/\xi_{\mathrm{s}}$, with $\xi_{\mathrm{s}}$ being the superconducting coherence length of the superconductor, may be used to estimate the strength of IPE\cite{Aarts1997}, although caution should be taken when making quantitative comparisons between different systems, as other factors, such as interfaces, can affect this ratio. In \rfig{f1}(d), we list $d_{\mathrm{cr,s}}/\xi_{\mathrm{s}}$ of our Al/V$_5$S$_8$ bilayer together with some S/N and superconductor/ferromagnet systems. In a Pb/Cu bilayer, $d_{\mathrm{cr,s}}/\xi_{\mathrm{s}}=0.025$\cite{Hilsch1962}, consistent with a weak IPE described by classical theories. $d_{\mathrm{cr,s}}/\xi_{\mathrm{s}}$ of ferromagnetic films can reach a pretty large value around 2\cite{Aarts1997,Strunk1994,Lazar2000,Muehge1997}. For our Al/V$_5$S$_8$ bilayer, $0.35<d_{\mathrm{cr,s}}/\xi_{\mathrm{s}}<0.73$ [calculated from \rfig{f1}(c)]. It is remarkable that the IPE of Al/V$_5$S$_8$ bilayer is even stronger than that of the Al/$(\mathrm{Co/Pd})_{10}$ film\cite{Xia2009}. Although V$_5$S$_8$ is certainly not a ferromagnet, antiferromagnetism cannot be completely ruled out, despite the absence of indirect evidence in electrical transport. In principle, the inverse proximity effect by an antiferromagnet should be much weaker than that by a ferromagnet, as the magnetic moments within the superconducting coherence length are compensated\cite{Bulaevskii2017,Huebener2002Sep}. However, magnetic defects can introduce pair-break scattering, leading to a strong suppression of superconducting transition temperature\cite{Davis1988}, which might explain our observation.

\begin{figure}[htbp]
	\begin{center}
		\includegraphics[width=1.0\columnwidth]{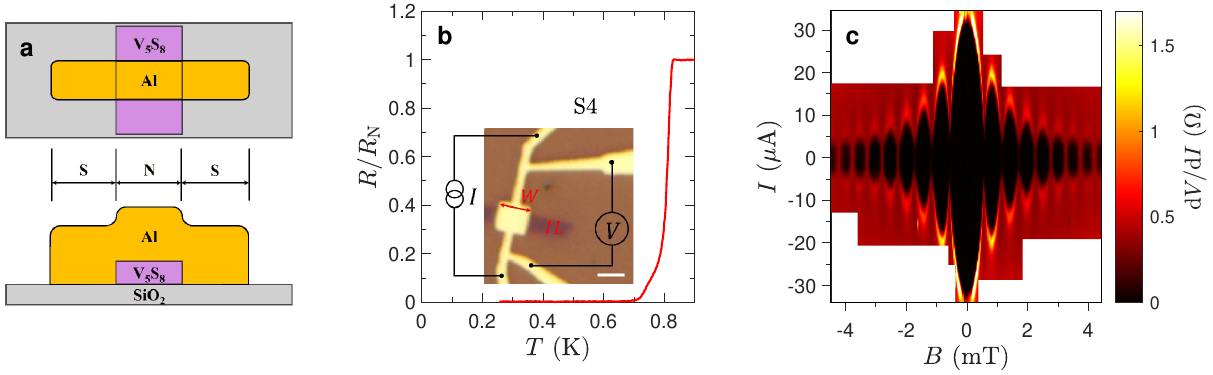}
		\caption{Homointerface planar JJ based on IPE. (a) The top view and the side view of the schematic of an Al{\textendash}(Al/V$_5$S$_8$){\textendash}Al JJ. (b) Normalized resistance $R$ as a function of $T$ for JJ S4. The inset is an optical micrograph of JJ S4 with an illustration of the measurement configuration. The junction length $L$ is 1.1 $\mu$m, the junction width $W$ is 2.7 $\mu$m. The scale bar is 2 $\mu$m. 0.6 $\mu$m away from the junction, the width of the Al superconductor is reduced to 0.8 $\mu$m so that no vortices can enter\cite{Stan2004}. Otherwise, flux jumping will appear in the Fraunhofer pattern. (c) Fraunhofer diffraction pattern of JJ S4 at 0.25 K.}
		\label{f2}
	\end{center}
\end{figure}

The IPE is intriguing and deserves further investigation. In this work, we focus on an application of the effect in JJs. Utilizing the observed strong IPE, an Al{\textendash}(Al/V$_5$S$_8$){\textendash}Al planar JJ can be fabricated using a simple one-step metal deposition process. As shown in \rfig{f2}(a), a narrow V$_5$S$_8$ flake and an Al strip form a crossbar. Non-superconducting Al on V$_5$S$_8$ plays the role of a weak link between superconducting Al electrodes on the SiO$_2$ substrate, which creates an SNS junction. Since the junction is built with a single piece of continuous Al film, the S/N homointerface is supposedly fully transparent. As Al(31)/V$_5$S$_8$(10) is non-superconducting at the lowest temperature of this experiment, the JJs studied below are all based on Al(31)/V$_5$S$_8$(10) bilayers. The temperature dependence of the resistance for JJ S4 shows that the Josephson supercurrent is established below 0.7 K. The differential resistance $\mathrm{d}V/\mathrm{d}I$ of the device displays clear diffraction patterns in the magnetic field vs bias current mapping [\rfig{f2}(c)]. In this Fraunhofer pattern, the characteristic of a JJ, at least eight sidelobes can be identified, suggesting a homogeneous junction with highly transparent interfaces. Note that similar JJs based on a cross structure have been demonstrated, using the IPE of ferromagnetic metals\cite{Vavra2013,Vavra2009}. However, the Fraunhofer pattern displays irregularities in amplitude and frequency. These features imply substantial inhomogeneities, probably stemming from grain or domain structures in the ferromagnetic layer. Our V$_5$S$_8$ samples are single crystals, enabling the formation of a uniform weak link.

The oscillating critical current $I_{\mathrm{c}}(B)$ can be described by $I_{\mathrm{c}}(B)=I_{\mathrm{c}}(0) \left| \frac{\sin(\pi BS/\Phi_0)}{\pi BS/\Phi_0}\right|$, where $B$ is the perpendicular magnetic field, $S$ is the effective area of junction, and $\Phi_0=h/2e$ is the flux quantum. Consistent with this relation, equally spaced nodes of the critical current can be seen in the pattern of JJ S4 [\rfig{f2}(c)]. The node spacing $\Delta B$ is 0.55 mT. $\Phi_0/\Delta B=3.8$ $\mu \mathrm{m}^2$ yields the effective area of the junction. In comparison, the nominal area of Al(31)/V$_5$S$_8$(10) is 3.0 $\mu \mathrm{m}^2$. The discrepancy can be ascribed to the London penetration depth and flux-focusing\cite{Suominen2017}.

\begin{figure}[htbp]
	\begin{center}
		\includegraphics[width=1\columnwidth]{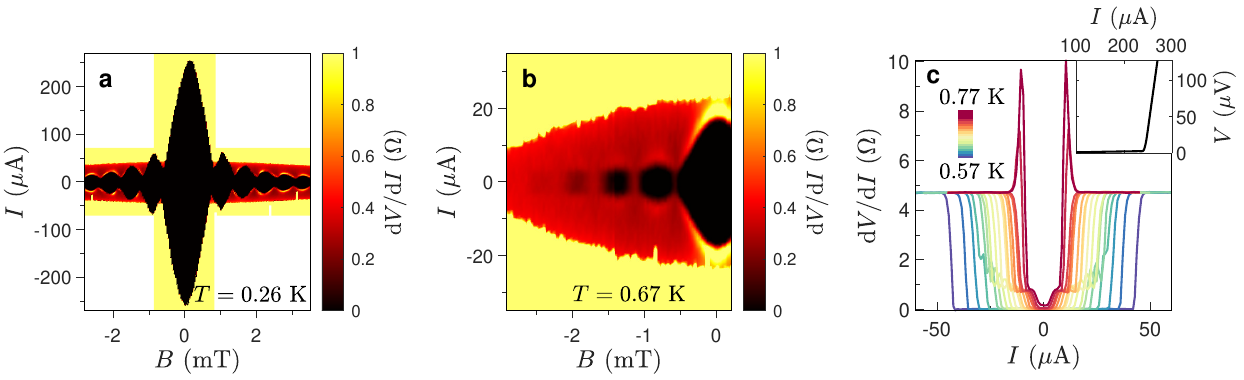}
		\caption{Properties of JJ S6. $L=0.9$ $\mu$m and $W=2.8$ $\mu$m. (a) and (b) Fraunhofer diffraction patterns at 0.26 and 0.67 K, respectively. (c) $\mathrm{d}V/\mathrm{d}I$ vs $I$ at different temperatures ranging from 0.57 to 0.77 K. The inset is the $I${\textendash}$V$ characteristic at 0.26 K.}
		\label{f3}
	\end{center}
\end{figure}

The good quality of the junction S/N interface helps the establishment of Josephson supercurrent across large gaps of the weak link. A zero resistance state and a finite supercurrent have been observed in JJ S5 with a $2.9$ $\mu$m gap, seen in the supplementary material. Likely for the same reason, our junction can support a giant supercurrent even when the gap is relatively large. \rfig{f3}(a) shows the Fraunhofer pattern of a device with a 0.9 $\mu$m gap. The regular and well-defined diffraction pattern confirms the uniformity of the junction. The supercurrent reaches 255 $\mu$A in the zero magnetic field at 0.26 K ($\sim 0.32 T_\mathrm{c}$), yielding a large supercurrent density of $2.1 \times 10^{5}$ $\mathrm{A/cm^{2}}$ among SNS junctions\cite{Lacquaniti2001,Abay2012,Frielinghaus2010}. The critical current of the junction is so large that the superconductivity of Al electrodes is quenched by the Joule heating as soon as the junction goes into the normal state, which will be explained shortly.

Because of the device structure, the measured junction resistance is a sum of the actual junctions resistance and the resistances of Al segments between two voltage probes, as seen in the inset of \rfig{f2}(b). Consequently, as the bias current increases, two sequential resistance transitions are expected. The first one indicates the critical current of the junction, while the second one is the critical current of Al electrodes, $I_\mathrm{c}^\mathrm{Al}$. This is what is observed above 1.5 mT, as seen in \rfig{f3}(a). The critical current line of Al encloses the Fraunhofer pattern. There are 0.4 $\Omega$ for the normal state resistance of junction and 4.3 $\Omega$ for the electrode resistance. However, below 1.5 mT, the zeroth and first diffraction peaks protrude over the Al critical current. In particular, the maximum of the zeroth peak exceeds the Al critical current by 4.8 times. This bizarre result seems to indicate that the Al electrode can sustain a much higher supercurrent when the junction is also in the superconducting state than that when the junction is non-superconducting. We believe that, rather than $I_\mathrm{c}^\mathrm{Al}$ being enhanced in the protruding regions, $I_\mathrm{c}^\mathrm{Al}$ is strongly suppressed in other regions because of the Joule heating occurring at the junction when it goes into the normal state. When the critical current of the junction is higher than the suppressed $I_\mathrm{c}^\mathrm{Al}$, two transitions take place simultaneously, leading to only one resistive transition in the protrusion regions. At a higher temperature, $T=0.67$ K, the critical current of the junction is reduced more strongly than $I_\mathrm{c}^\mathrm{Al}$. The whole Fraunhofer pattern submerges below $I_\mathrm{c}^\mathrm{Al}$ and a common two-transition pattern appears, as shown in \rfig{f3}(b).

\begin{figure}[htbp]
	\begin{center}
		\includegraphics[width=0.5\columnwidth]{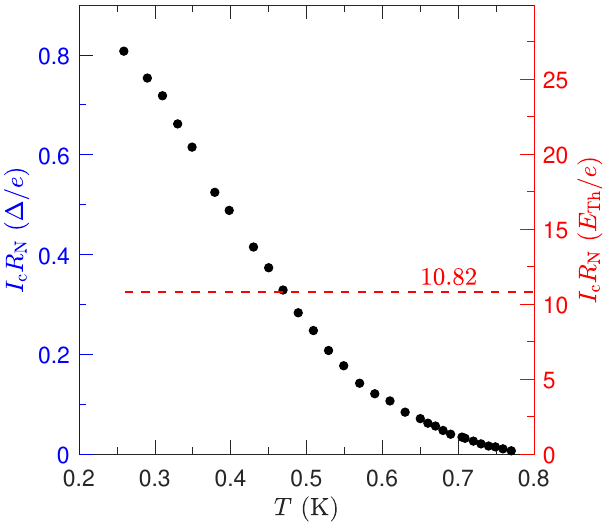}
		\caption{$I_\mathrm{c} R_\mathrm{N}$ of JJ S6 scaled by $\Delta$ and $E_\mathrm{Th}$ as a function of temperature. The red dashed line indicates the theoretical value of $10.82E_\mathrm{Th}/e$\cite{Dubos2001}.}
		\label{f4}
	\end{center}
\end{figure}

Transparent S/N interfaces in SNS JJs are critical for obtaining a large $I_\mathrm{c} R_\mathrm{N}$, which is an important parameter of JJs. Shabani \textit{et al.} improved the interface by employing epitaxial growth of aluminum on a semiconductor and achieved $I_\mathrm{c} R_\mathrm{N}\sim0.68\Delta/e$ at very low temperature, $T/T_\mathrm{c}=0.02$\cite{Shabani2016}. Here, $\Delta$ is the superconducting gap of the superconductor. We plot $I_\mathrm{c} R_\mathrm{N}$ of JJ S6 as a function of temperature in \rfig{f4}. At 0.26 K($T/T_\mathrm{c}=0.32$), $I_\mathrm{c} R_\mathrm{N} \approx 0.81\Delta/e$ is obtained. This value is anticipated to be substantially enhanced with decreasing temperature\cite{Dubos2001}. Generally speaking, $I_\mathrm{c} R_\mathrm{N}$ is bounded by the minimum of $\Delta$ and the Thouless energy $E_\mathrm{Th}$. In the diffusive limit, $E_\mathrm{Th}$ is given by $\hbar D/L^2$, with $D$ being the diffusion constant and $L$ being the junction length\cite{Dubos2001}, while $E_\mathrm{Th}$ is $\hbar v_\mathrm{F}/L$ in the ballistic limit, with $v_\mathrm{F}$ being the Fermi velocity\cite{Argaman1999May,BenShalom2016}. $D$ can be calculated by $D=\frac{1}{3}(\frac{\pi k_\mathrm{B}}{e})^2 \frac{\sigma}{\gamma}$, where $\gamma$ is the electronic specific heat coefficient and $\sigma$ is the electrical conductivity\cite{Pippard1960}. Since the electrical conductivity of the Al film is one order of magnitude larger than that of V$_5$S$_8$, $D$ of the Al/V$_5$S$_8$ bilayer is predominantly determined by the Al layer. Taking $\gamma=1.4 \times 10^{2}$ J$\cdot$m$^{-3}\cdot$K$^{-2}$ for Al\cite{Kittel2005} and $\sigma=2.6 \times 10^7$ S$\cdot$m$^{-1}$ for the weak link in JJ S6, $D$ turns out to be 4.5$\times 10^{-3}$ m$^2 \cdot$s$^{-1}$.  Through the relation $D=v_\mathrm{F} l_\mathrm{e}/3$ and $v_\mathrm{F}=2.02\times 10^8$ cm$\cdot$s$^{-1}$\cite{Kittel2005}, the mean free path $l_\mathrm{e}$ is 6.7 nm, much less than $L=0.9$ $\mu$m, which indicates that the junction is in the diffusive limit. Then, $E_\mathrm{Th}$ of JJ S6 becomes 3.7 $\mu$eV, much less than $\Delta$. In this long-junction limit, detailed theoretical calculations showed that for fully transparent interfaces, $I_\mathrm{c} R_\mathrm{N}$ at zero temperature is $10.82E_\mathrm{Th}/e$\cite{Dubos2001}. Surprisingly, our $I_\mathrm{c} R_\mathrm{N}\sim 26.9E_\mathrm{Th}/e$ at an intermediate temperature of 0.26 K, already significantly higher than the theoretical value.

Several artifacts may lead to an $I_\mathrm{c} R_\mathrm{N}$ larger than the theoretical calculation. First, nonuniformity, for instance, a path of superconductor, can sustain a large supercurrent. However, the regular Faunhofer pattern indicates that the junction is uniform. Second, if the actual junction length $L$ is shorter than the nominal length $0.9$ $\mu$m, $E_\mathrm{Th}$, inversely proportional to $L^2$, will be larger. On the contrary, the junction length estimated from the period of the Fraunhofer pattern is actually larger than $0.9$ $\mu$m. The extra length can be accounted for by screening and flux-focusing. Thus, the real length is close to the nominal one. Third, the calculation of $D$ assumes that the carrier density and the band structure of Al on V$_5$S$_8$ remain the same as those of bulk Al. This may not be true when the film is ultra-thin, as a Coulomb gap is known to develop\cite{Altshuler1985}, but our Al film is too thick to be in that regime. From $E_\mathrm{Th}=\hbar D/L^2$, one can see that the extraordinary $I_\mathrm{c} R_\mathrm{N}$ is consistent with the observed long-range Josephson coupling. We believe that both are related to the fact that the weak link is made from a superconductor that is forced to stay in its normal state by the inverse proximity effect. Though non-superconducting, the pairing interaction remains, which helps establish an unexpectedly strong and long-range Josephson coupling. An enhancement of the proximity effect because of the pairing interaction has recently been revealed in a scanning tunneling microscopy study\cite{Cherkez2014Mar}.

The high transparency of the S/N interface of our JJs is also reflected in the excess current, defined by $I_{\mathrm{ex}}=I-V/R_\mathrm{N}$. $I_{\mathrm{ex}}$ can be obtained by extrapolating the linear dependence of the $I${\textendash}$V$ characteristic in the normal state to the $I$ axis. The higher the interface transparency, the higher probability the Andreev reflection occurs at, hence the larger $I_{\mathrm{ex}}$\cite{Blonder1982}. The inset of \rfig{f3}(c) shows $I_{\mathrm{ex}} \approx I_\mathrm{c}$, implying highly transparent interfaces.

\section{CONCLUSION}

In conclusion, we observed a strong IPE in a bilayer of Al/V$_5$S$_8$. Based on the effect, a Josephson junction with superconductor{\textendash}normal metal homointerface can be readily fabricated. Owing to the highly transparent interface, the junction displays a large critical current and $I_\mathrm{c} R_\mathrm{N}$ product. Moreover, the homointerface is beneficial for large-scale fabrication of consistent devices. Therefore, our method shows potentials in superconducting electronic applications.

\section*{SUPPLEMENTARY MATERIAL}

See the supplementary material for the properties of $\mathrm{V_5 S_8}$, the detailed calculation of the proximity effect in the Al/$\mathrm{V_5 S_8}$ bilayer, the measurement of (Al/V$_5$S$_8$){\textendash}V$_5$S$_8${\textendash}(Al/V$_5$S$_8$) Josephson junction, and the measurement of Al{\textendash}(Al/V$_5$S$_8$){\textendash}Al Josephson junction S5.

\begin{acknowledgements}
We are grateful for helpful discussions with J. Linder and Q. F. Sun. This work was supported by National Key Basic Research Program of China (No. 2020YFA0308800) and NSFC (Project Nos.
11774009 and 12074009).
\end{acknowledgements}

\section*{DATA AVAILABILITY}
The data that support the findings of this study are available from the corresponding author upon reasonable request.

%

\clearpage

\pagebreak
\widetext
\begin{center}
\textbf{\large Supplemental Materials}
\end{center}
\setcounter{equation}{0}
\setcounter{figure}{0}
\setcounter{table}{0}
\setcounter{page}{1}
\makeatletter
\renewcommand{\theequation}{S\arabic{equation}}
\renewcommand{\thefigure}{S\arabic{figure}}
\renewcommand{\bibnumfmt}[1]{[S#1]}
\renewcommand{\citenumfont}[1]{S#1}

This Supplemental Material Section contains the properties of $\mathrm{V_5 S_8}$, detailed calculation of the proximity effect in the Al/$\mathrm{V_5 S_8}$ bilayer, the measurement of (Al/V$_5$S$_8$){\textendash}V$_5$S$_8${\textendash}(Al/V$_5$S$_8$) Josephson junction and Al{\textendash}(Al/V$_5$S$_8$){\textendash}Al Josephson junction S5.

\subsection{Crystal structure of superlattice V$_5$S$_8$}

\begin{figure}[htp]
\includegraphics[width=0.35\textwidth]{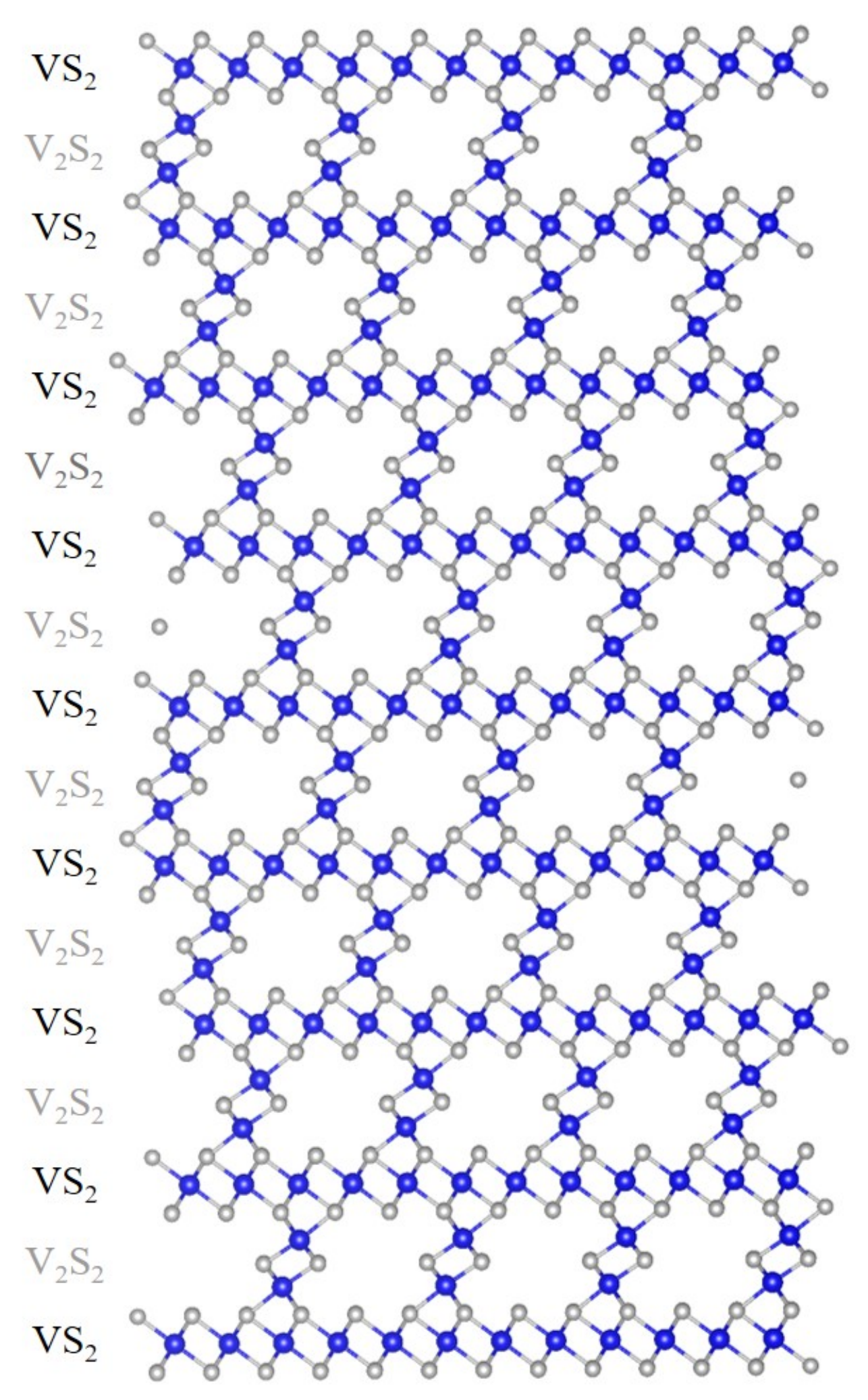}
\caption{Atomic structure of the V$_5$S$_8$ superlattice in side view.}
\label{Fig_S1}
\end{figure}

The crystal structure of the superlattice $\mathrm{V_5 S_8}$ has a triclinic symmetry and belongs to the space group of P1, with lattice constants of $a=9.69$ \AA, $b=3.23$ \AA, $c=75.53$ \AA, $\alpha=\beta=90^\circ$ and $\gamma=120^\circ$. \rfig{Fig_S1} depicts the atomic structure of the $\mathrm{V_5 S_8}$ superlattice in side view. One can see that $\mathrm{V S_2}$ layers are intercalated with $\mathrm{V_2 S_2}$ atomic chains, which are oriented perpendicular to the plane of paper.

\FloatBarrier

\subsection{Hall resistance and temperature dependence of the resistance of V$_5$S$_8$}

\begin{figure}[htp]
\includegraphics[width=1.0\textwidth]{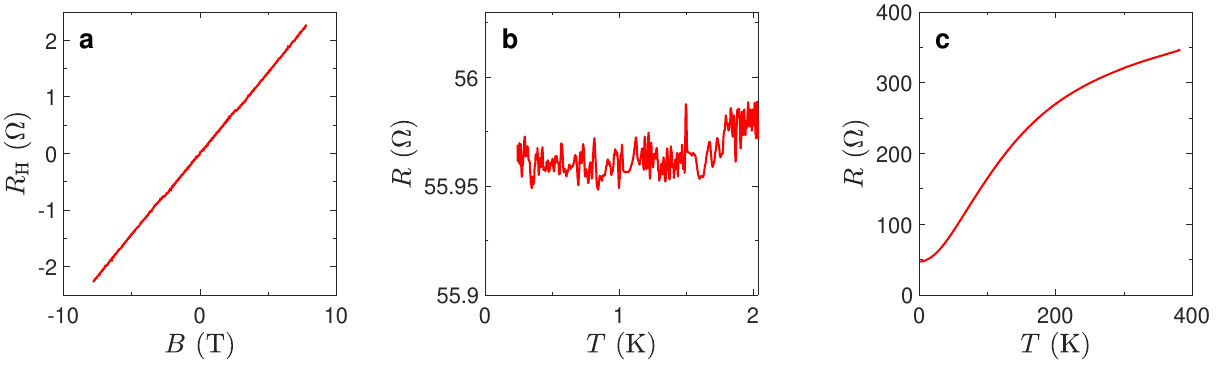}
\caption{Hall and longitudinal resistances of V$_5$S$_8$. (a) Magnetic field dependence of Hall resistance at 0.24 K. (b) Temperature dependence of the resistance from 0.24 to 2 K. (c) Temperature dependence of the resistance from 2 to 380 K of another V$_5$S$_8$ sample.}
\label{Fig_S4}
\end{figure}

As shown in \rfig{Fig_S4}(a), the Hall resistance of V$_5$S$_8$ is linear in magnetic field at 0.24 K. Absence of an anomalous Hall effect that is nonlinear in magnetic field indicates no ferromagnetic order. The temperature dependent resistivity of antiferromagnetic metals often displays a kink at the N\'{e}el temperature. However, the resistance curves of our V$_5$S$_8$ from 0.24 to 380 K are smooth and do not have any kink[\rfigs{Fig_S4}(b) and S2(c)].

\FloatBarrier

\subsection{Calculation of the proximity effect based on the Werthamer theory}

\begin{figure}[htp]
\includegraphics[width=0.8\textwidth]{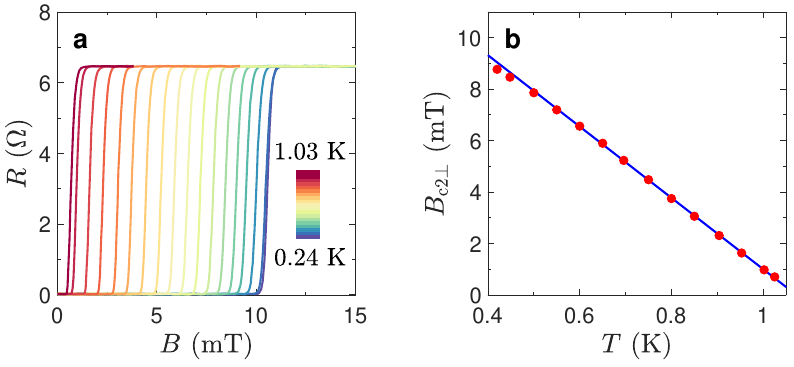}
\caption{Magnetic field dependence of resistance of a 72-nm-thick Al film on the SiO$_2$ substrate. (a) $R${\textendash}$B$ curves at different temperatures. (b) The linear-$T$ dependence of $B_\mathrm{c2\perp}$.}
\label{Fig_S5}
\end{figure}

According to the theory by Werthamer, the proximity effect in the one-dimension S/N bilayer can be described by a set of three equations\cite{Werthamer1963a,Hauser1964a,Nagel1994a}:
\begin{eqnarray}
-\chi(-\xi_{\mathrm{n}}^2 k_{\mathrm{n}}^2) & = & \ln(T_{\mathrm{c}}/T_{\mathrm{cn}})
\label{1} \\
\chi(\xi_{\mathrm{s}}^2 k_{\mathrm{s}}^2) & = & \ln(T_{\mathrm{cs}}/T_{\mathrm{c}})
\label{2} \\
\left[N\xi^2 k\tan(kd)\right]_{\mathrm{s}} & = & \left[N\xi^2 k\tanh(kd)\right]_{\mathrm{n}}
\label{3}
\end{eqnarray}
. Here $T_{\mathrm{c}}$ is the transition temperature of the S/N bilayer, $T_{\mathrm{cs}}$ and $T_{\mathrm{cn}}$ are the transition temperatures of the superconductor and the normal metal, respectively. $\xi_{\mathrm{s}}$ is the superconducting coherence length of the superconductor and $\xi_{\mathrm{n}}$ is the depth by which Cooper pairs penetrate into the normal metal. $\chi(Z)=\psi(\frac{1}{2}+\frac{1}{2}Z)-\psi(\frac{1}{2})$, where $\psi$ is the digamma function. $k_{\mathrm{s,n}}$ are free parameters, $N$ is the density of state, $d$ is the thickness. \reqs{1} and (\ref{2}) describe the properties of normal metal and superconductor, respectively. \req{3} is the boundary condition at the S/N interface.

Since $\mathrm{V_5 S_8}$ is a normal metal, $T_{\mathrm{cn}}=0$. Plugging it into \req{1}, we get $-\chi(-\xi_{\mathrm{n}}^2 k_{\mathrm{n}}^2)=+\infty$, so $k_{\mathrm{n}}=1/\xi_{\mathrm{n}}$. Using this relation, \req{3} becomes
\begin{equation}
[N\xi^2 k\tan(kd)]_{\mathrm{s}}=[N\xi\tanh(d/\xi)]_{\mathrm{n}}
\label{4}
\end{equation}
. For a nonmagnetic diffusive system, $\xi_{\mathrm{n}}=(\hbar D_{\mathrm{n}}/2 \pi k_{\mathrm{B}}T)^{1/2}$, where $D_{\mathrm{n}}$ is the diffusion coefficient\cite{Eschrig2015}. Generally, $\xi_{\mathrm{n}}$ is much larger than $d_{\mathrm{n}}=10$ nm\cite{Keizer2006}, so \req{4} can be approximated to
\begin{equation}
[N\xi^2 k\tan(kd)]_{\mathrm{s}}=[Nd]_{\mathrm{n}}
\label{5}
\end{equation}
. Assuming $\mathrm{V}_5 \mathrm{S}_8$ and Al as free electron systems, we have $N_{\mathrm{n}}/N_{\mathrm{s}}=(n_{\mathrm{n}}/n_{\mathrm{s}})^{1/3}$, where $n$ is the carrier density. For Al, $n_{\mathrm{s}}=1.806$ $\times 10^{29}$ $\mathrm{m^{-3}}$, while for $\mathrm{V}_5 \mathrm{S}_8$, $n_{\mathrm{n}}=4.161$ $\times 10^{27}$ $\mathrm{m^{-3}}$, determined from the Hall resistivity\cite{Zhou2021a}. Then, $N_{\mathrm{n}}/N_{\mathrm{s}}$ is equal to 0.2846. According to the Ginzburg-Landau theory, the perpendicular upper critical field $B_{\mathrm{c2}\perp}$ of the superconducting film is linear with the temperature $T$ as $B_{\mathrm{c2}\perp}(T)=B_{\mathrm{c2}\perp}(0)(1-T/T_{\mathrm{cs}})$ near $T_{\mathrm{cs}}$. $\xi_{\mathrm{s}}$ is related to $B_{\mathrm{c2}\perp}(0)$ via $\xi_{\mathrm{s}}=\frac{1}{\pi} (\frac{2 \Phi_{0}}{\pi B_{\mathrm{c2}\perp}(0)})^{1/2}$, where $\Phi_0=h/2e$ is the flux quantum\cite{Lazar2000a}. Therefore, by measuring the magnetic field dependence of resistance of Al film depicted in \rfig{Fig_S5}, we find that $\xi_{\mathrm{s}}$ is about 102 nm. Finally, \req{5} is reduced to
\begin{equation}
[k\tan(kd)]_{\mathrm{s}}=2.846 \times 10^{-4}\;\mathrm{nm^{-1}}.
\label{6}
\end{equation}

\FloatBarrier

\subsection{Measurement of (Al/V$_5$S$_8$){\textendash}V$_5$S$_8${\textendash}(Al/V$_5$S$_8$) Josephson junction}

\begin{figure}[htp]
\includegraphics[width=0.8\textwidth]{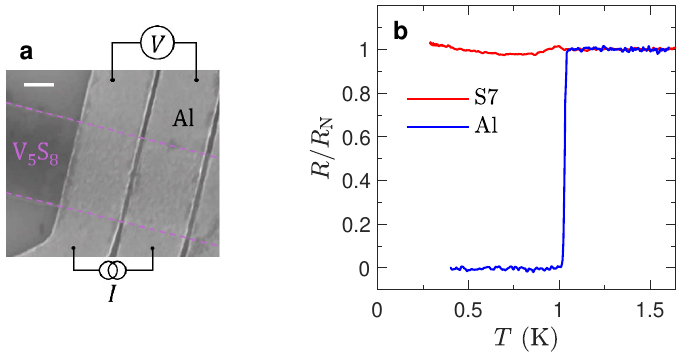}
\caption{Properties of (Al/V$_5$S$_8$){\textendash}V$_5$S$_8${\textendash}(Al/V$_5$S$_8$) Josephson junction S7. (a) An SEM micrograph of the junction with an illustration of the measurement configuration. The scale bar represents 400 nm. (b) Normalized resistance $R$ as a function of temperature $T$ for the junction and the 46-nm-thick Al film on SiO$_2$.}
\label{Fig_S6}
\end{figure}

We have attempted to fabricate a couple of conventional SNS Josephson junctions by placing closely-spaced Al electrodes on V$_5$S$_8$. As shown in \rfig{Fig_S6}(a), two parallel 46-nm-thick Al strips span over a narrow 10-nm-thick V$_5$S$_8$ flake. The gap between two Al strips ranges from 74 to 400 nm. The 46-nm-thick Al strip on SiO$_2$ goes to the superconducting state at 1.03 K, whereas no Josephson supercurrent was ever observed in all these devices down to 0.29 K [\rfig{Fig_S6}(b)]. The absence of supercurrent in junction with an extremely narrow gap implies a strong inverse proximity effect of V$_5$S$_8$.

\FloatBarrier

\subsection{Measurement of Al{\textendash}(Al/V$_5$S$_8$){\textendash}Al Josephson junction S5}

\begin{figure}[htp]
\includegraphics[width=0.8\textwidth]{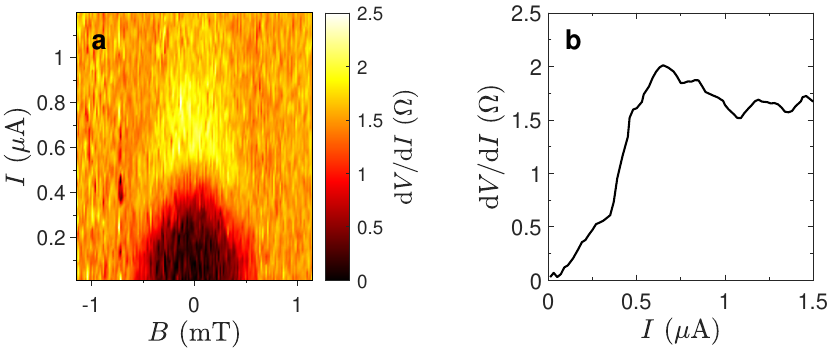}
\caption{Properties of Josephson junction S5. The junction length $L=2.9$ $\mu$m, the junction width $W=1.6$ $\mu$m. (a) Two dimensional mapping of the differential resistance $\mathrm{d}V/\mathrm{d}I$ in the $I${\textendash}$B$ plane at 0.26 K. (b) $\mathrm{d}V/\mathrm{d}I$ versus $I$ at $B=0$ extracted from (a).}
\label{Fig_S3}
\end{figure}

\FloatBarrier

\clearpage

\end{document}